\documentclass[fleqn,twoside]{article}
\usepackage{espcrc2}

\usepackage{graphicx}
\usepackage[figuresright]{rotating}

    \newcommand{\gr}{$\gamma$-ray \,}
    \newcommand{\grs}{$\gamma$-rays \,}

\newcommand{\AmS}{{\protect\the\textfont2
  A\kern-.1667em\lower.5ex\hbox{M}\kern-.125emS}}

\title{High Energy Gamma-Ray Astronomy}

\author{H.J. V\"olk\address{Division of Astrophysics, 
        Max Planck Institute for Nuclear Physics, \\ 
        P.O. Box 103980, 69029 Heidelberg, Germany}%
        }
       
\begin{document}

\begin{abstract} 
The physics results of high energy \gr astronomy are reported, emphasizing
recent achievements with ground-based detectors. This includes some of the
instrumental developments and latest projects. The fundamental contribution of
the field to the question of Cosmic Ray origin is highlighted.
\vspace{1pc} 
\end{abstract}

\maketitle

\section{PHYSICS GOALS}

\noindent High energy \gr astronomy is connected with three important
areas of basic science:

\begin{itemize}
\item High Energy Astrophysics
\item Observational Cosmology
\item Astroparticle Physics
\end{itemize}

\noindent Overall and in somewhat simplified terms, {\bf High Energy
Astrophysics} concerns itself with the most energetic and violent
processes in the Universe. High energy \gr astronomy can specifically
study the {\it nonthermal} processes and in an global sense its field is
the {\it Nonthermal Universe}. Having said this, the most interesting
objects are those which produce the nonthermal particle populations. I
will mention here {\it Pulsars} and {\it Supernovae}, and possibly {\it
new types of sources}. The long list of other Galactic topics of interest
contains accreting X-ray binaries ($\mu$-Quasars), the diffuse Galactic
emission, molecular clouds, etc. For a recent review see \cite{Kifu03}.
The main extragalactic \gr detections \cite{Week03} are for supermassive
Black Holes as active galactic nuclei (AGNs) in the centers of galaxies.
They have different forms of appearence, from giant {\it Radio Galaxies}
and {\it Blazars} to {\it BL Lac objects} with their relativistic jets
pointing directly towards us. Possibly a nearby {\it Starburst Galaxy} has
been detected in very high energy \grs \cite{Itoh02}. I shall leave aside
here Gamma Ray Bursts, luminous Quasars or the Extragalactic \gr
Background and rather come back to them in the last section.

\begin{figure}[ht]
\centering
\vspace*{-1cm}
\includegraphics[width=7.5cm]{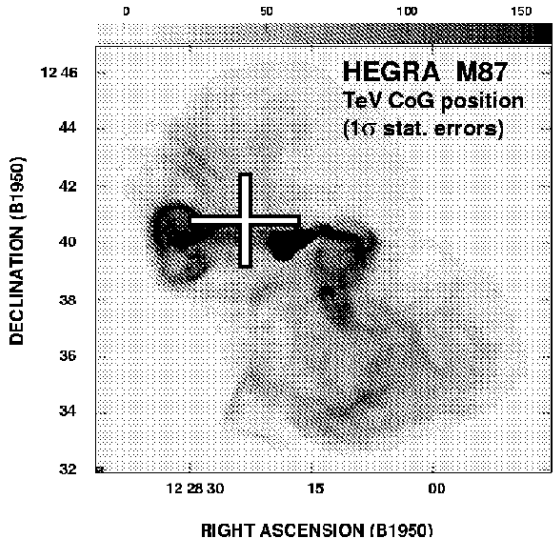}
\caption{Radio image of M87 and its jet in the Virgo cluster at 90 
cm. The cross (statistical $1\sigma$ errors) marks the position of the TeV 
\gr excess detected by HEGRA \cite{Aha03b}.} \label{f1}
\end{figure}  

The role of \gr astronomy in {\bf Observational Cosmology} concerns cosmic
structure formation in terms of stars, galaxies and clusters of galaxies.  
One topic is the diffuse extragalactic radiation field (CIB) from these
objects in the optical/infrared wavelength range. It is not easily measured
directly due to dominant foregrounds, from Zodiacal Light to Galactic
Cirrus. Pair production of high energy \grs on the CIB photons
however leads to characteristic absorption features in the \gr
spectrum of distant sources like AGNs. For given distance and \gr source
spectrum this determines the CIB spectrum. Including the intergalactic
magnetic field, the CIB absorption/CMB Inverse Compton cascade leads to
Giant Pair Halos around sources with sufficiently energetic primary \gr
photon emission $\gg 10$~TeV \cite{Ahar94}, whose angular structure and
spectrum in principle allow the derivation of CIB spectrum - locally in
redshift - as well as absolute distance. This is the \gr analog to the
Sunyaev-Zel'dovich effect \cite{Suny72} on the CMB in clusters of galaxies
and shows the richness of high-energy photon astronomy. A topic for the
near future concerns the study of formation and internal evolution of
galaxy clusters from their characteristic \gr spectrum and morphology (e.g.
\cite{Voel96,Mini03,Gabi03}).

In {\bf Astroparticle Physics} $\geq 30$~TeV \gr observations permit an
indirect Dark Matter search through the possible detection of generic SUSY
WIMP annihilation features in Dark Matter Halos like the one expected in the
center of our Galaxy \cite{Berg03}. It is not clear whether this will
be possible in practice because of a number of astronomical backgrounds,
even if such annihilation occurs. However the result would be complementary
to a detection of supersymmetric particles in accelerators like the LHC and
would go a long way towards the identification of the nonbaryonic Dark
Matter constituent(s) in the Universe. \gr observations of the Galactic
Center are therefore high on the list of all \gr instruments that can reach
this point in the sky.

\subsection{Introduction}

In many respects high energy \gr astronomy is a young field. Major results
have been obtained since little more than a decade, and a new generation
of mature instruments is only now starting operations. After a short
description of the instruments I will mainly discuss results obtained with
ground-based detectors. This emphasis justifies itself simply by the more
recent developments in that area. The satellite-based instruments and
their achievements have already been reviewed extensively in the past
(e.g. \cite{Scho01}). 

The talk will be concerned with some of the nonthermal sources mentioned
above. Of the diffuse nonthermal particle populations I will discuss the
question of Cosmic Ray origin. The dynamically dominant part of this
population extends to the "knee" region at a few $\times 10^{15}$~eV and
contains essentially all the energy density in Cosmic Rays. I will indicate
that \gr astronomy in conjunction with X-ray astronomy and theory is closing
in on a physics solution in terms of supernova explosions and their remnants
(SNRs). I think that the remaining uncertainties are primarily due to poor
\gr detection statistics rather than incomplete theoretical understanding.

Beyond the present instruments fascinating developments suggest
themselves. I will summarize some of them in the last section.
\begin{figure*}[t]
\centering
\includegraphics[width=15cm]{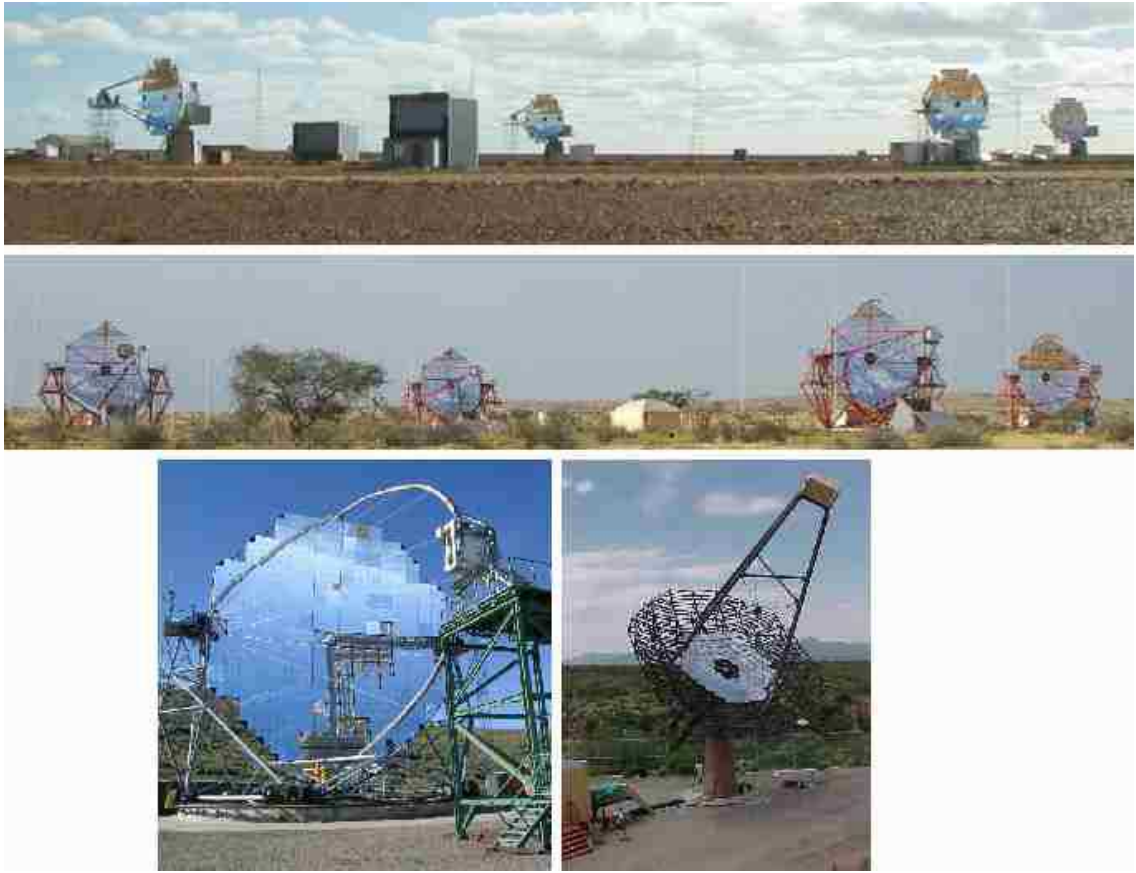}
\caption{The 4-telescope array CANGAROO-III (top). The 4 H.E.S.S. I
telescopes (middle). MAGIC telescope (bottom left). VERITAS prototype
(bottom right).} 
\label{f2} 
\end{figure*}

\section{GAMMA-RAY INSTRUMENTS}

In the past the field was dominated by the suite of {\it satellite-based}
spark chamber detectors with charged particle anticoincidence shields
SAS~II~(1972) $\Rightarrow$~ COS~B~(1975) $\Rightarrow$~EGRET~(1991),
sensitive in the range $30~{\mathrm MeV} < E_{\gamma} < {\mathrm
{few}}$~tens of GeV. Despite their large FoV of several steradian, life
time of several years, and reasonable energy resolution $\Delta E/E \sim
20$~\%, the low angular resolution $\geq 1^{\circ}$ has largely prevented
the identification of compact sources. The EGRET instrument found about
300 sources \cite{Hart99}, many of them AGNs, but the nature of the
majority is still unknown. The logical consequence was the GLAST project,
to be launched in late 2006, and its smaller Italian brother AGILE (2004),
based on solid state converter/tracker technology and an angular
resolution of $\sim 1/10$ of a degree. GLAST, with an effective detector
area $\sim 1 m^2$, can detect \grs up to hundreds of GeV. In practice,
statistics will limit the energy range to a few tens of GeV.

The detection area, which is intrinsically small for space detectors, reaches
very large values $\geq 10^4 {\mathrm m}^2$ for {\it ground-based detectors}
in the form of imaging atmospheric Cherenkov telescopes (IACTs). Their energy
thresholds are about 100 GeV, with $\Delta E/E \leq 20$~\% and an angular
resolution of $10^{-1}$~degree. Disadvantages are the low observation
efficiency $\leq 10$~\% and the intrinsically small FoV $\sim 
10^{-4}$~sr.  
About a dozen sources have been found until summer 2003, comparatively
well-studied, one of them unidentified.

Astronomical results are mainly from the first generation (Whipple,
$10$~m) and second generation (CANGAROO-I, $3.8$~m; HEGRA, $5\times
3.3$~m; CAT $\sim 5$~m) telescopes. But a third generation of major
experiments with 10-fold sensitivity at 1 TeV and energy threshold of 100
GeV or below is beginning to take data at various stages of completion.
They are CANGAROO-III (2001, $4 \times 10$~m) in Australia, H.E.S.S. I
(2002; $4\times 12$~m) in Namibia, MAGIC (2003, $17$~m) on La Palma, and
VERITAS (2003; $4-7\times 12$~m) on Kitt Peak, Arizona. HEGRA pioneered a
stereoscopic array of 5 telescopes operating in coincidence. CANGAROO-III,
H.E.S.S. and VERITAS took over this technique, and MAGIC also plans to
build a 2nd and possibly a 3rd identical telescope.

\section {PHYSICS RESULTS}

As mentioned before, I will only discuss the most recent results, from 
2002 onwards. For general overviews, see \cite{Kifu03,Week03}.
\begin{figure}
\centering
\includegraphics[width=7.5cm]{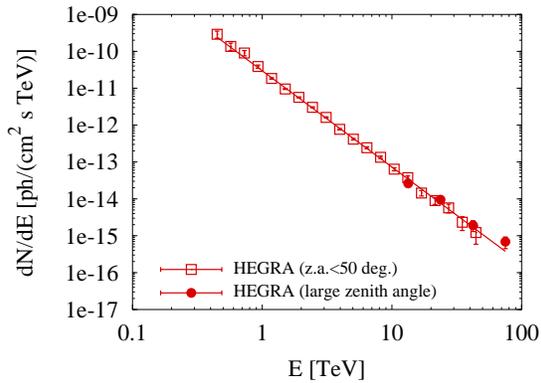}
\caption{HEGRA differential photon spectrum of the Crab Nebula in the TeV 
range \cite{Horn03}.}
\label{f3}
\end{figure}


The {\bf Crab Nebula} is the strongest steady source in the Northern Sky
and has assumed the role of a standard candle. Its emission is dominantly
nonthermal, presumably from particles accelerated in the termination shock
of a pair plasma wind with a bulk Lorenz factor of about $10^7$ that
carries away the rotational energy loss of the neutron star. Despite its
very large magnetic field of $300~\mu$G the emission is dominated by
synchrotron radiation.  High energy \grs had been assumed to be the result
of inverse Compton (IC) emission by the synchrotron electrons. Yet their
energy spectrum has been recently shown to extend up to at least 70 TeV,
unexpected from IC modeling because of the Klein-Nishina effect
{\cite{Horn03}). An interesting question, also relevant to the
acceleration process as such, is whether these very high energy photons
come from a hadronic component in the wind with a hard spectrum.


With the HEGRA stereoscopic array it was possible to make a short-exposure
scan of the Galactic Plane \cite{Aha01a} in search for new Galactic
sources.  Only upper limits were found. However, by source stacking a
combined upper limit at twice the theoretically predicted total
$\pi^0$-decay flux could be set for the 19 known shell-type SNR candidate
sources from the {\bf Galactic SNR population} in the search field. This
is still consistent with expectations and indicates a dominant hadronic
emission. I will discuss individual SNRs in the last section.
\begin{figure}[t]
\centering
\vspace*{-1cm}
\includegraphics[width=7cm]{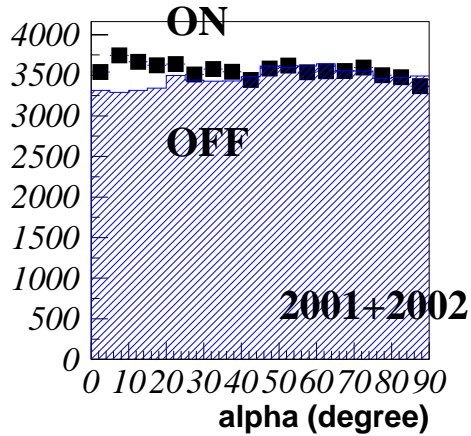}
\caption{CANGAROO directional (``alpha'') ON-source ({\it squares}) and 
OFF-source
({\it hatched}) distributions for the Galactic Center \cite{Tsuc03}.} 
\label{f4}
\end{figure}
The single CANGAROO-II telescope in Australia (a precursor to
CANGAROO-III) has found a signal from the {\bf Galactic Center} at 400 GeV
\cite{Tsuc03}, thereby confirming the EGRET detection in the GeV range.
The combined significance from observations in the two years 2001 and 2002
is approaching $10~\sigma$ (Fig. 4). The detection has raised high 
expectations for
the results with the full CANGAROO-III and H.E.S.S. I arrays in the
Southern Henisphere that will start operations in 2004. What is needed is
a TeV-spectrum and morphological information.
\begin{figure}
\centering
\vspace*{-2cm}
\includegraphics[width=7.5cm]{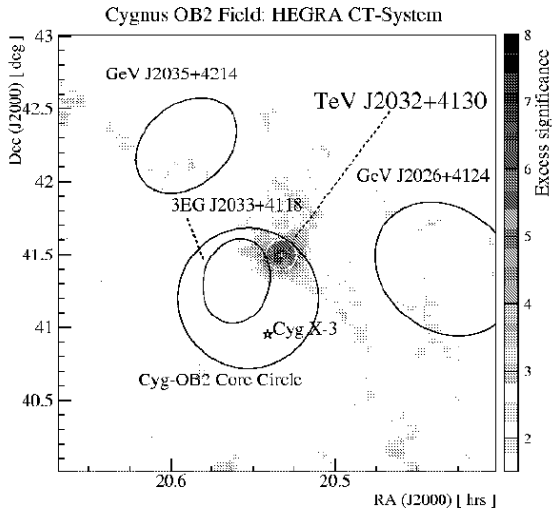}
\caption{TeV significance map of the Cygnus OB2 field, including GeV error 
ellipses and positions of known sources \cite{Aha02a}.} \label{f5}
\end{figure}


In a serendipitous discovery, the HEGRA array has found an {\bf
Unidentified TeV source} at a level of $7~\sigma$ in the Cygnus region of
the Galactic disk, just $0.5^{\circ}$ north of Cygnus X-3 \cite{Aha02a}.
The source has no counterpart in any other wavelength range and is
marginally extended. The photon spectrum exhibits a hard power law ($dN/dE
\propto E^{-1.9}$). From all this evidence, or lack of, the source should
be in fact hadronic. Previously it was thought that sources in the TeV
region would be associated with astronomical objects of such brightness
that they would be well visible at lower energies. However this seems not
to be the case. The source is located in the Cygnus OB2 association, one
of the most active regions of the Galaxy, with many massive stars.
Interactions of their rarefied fast winds might be able to produce such a
unique signal \cite{Butt03}.


{\bf Active Galactic Nuclei (AGNs)} in the form of BL Lac objects have
their relativistic jets pointing towards the observer. The Doppler
boosting makes these jets obvious candidates for \gr emission. BL Lacs,
and more generally Blazars, are also known to exhibit strong time
variations down to fractions of an hour, which may give important clues
for the nature of the jet and are detected with TeV instruments but not
with the GeV detectors in space. On the other hand very luminous Quasars
have not been seen above the GeV range, probably due to internal high
energy \gr absorption. Thus, for the family of AGNs at large, the 
GeV-range
and the TeV range are complementary. The cosmological aspect of the TeV
sources is their absorption on the CIB.  The best example found until is
the Blazar H1426+428 at redshift $z\approx 0.13$. Fig. 6 applies the
absorption features from three models of the CIB to the data that show a
generic hardening above 1 TeV. The de-absorbed source spectra
\cite{Aha03a} are only physically "reasonable" for the case a). One task
for the new instruments is to find sources at higher $z$ in order to
further strengthen these results.
\begin{figure*} 
\centering 
\includegraphics[width=15cm]{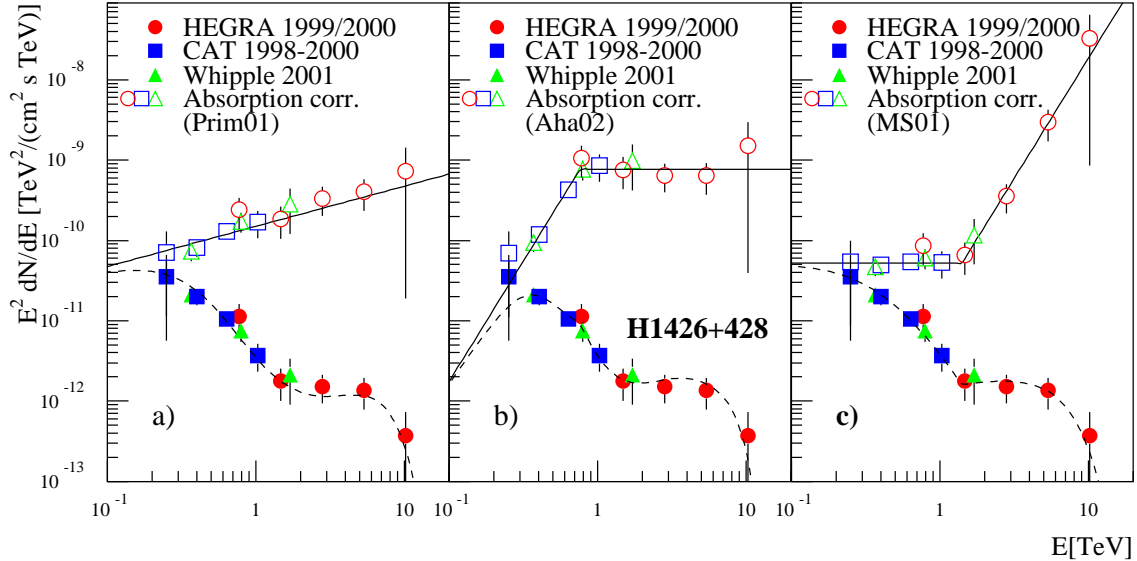}
\caption{Differential TeV energy spectrum of H1426+428, combining CAT,
Whipple and HEGRA data, and de-absorbed for three different CIB models. 
Absorbed broken power law fits ({\it solid lines}) are 
given by the 
{\it dashed curves} \cite{Aha03a}.} \label{f6} 
\end{figure*}

{\bf Radio Galaxies} show jets at oblique angles. Therefore little Doppler
favoritism is expected. It was therefore rather unexpected when HEGRA
indeed detected a weak signal from M87 (Fig. 1) in the center of the Virgo
cluster at a distance $\leq 20$~Mpc \cite{Aha03b}. The obvious question
is whether this shows a new class of extragalactic sources from the AGN
family, or whether we have here a first cluster source coming about
through the expected confinement of energetic particles in rich clusters
over periods that exceed a Hubble time \cite{Pfro03}. The question remains
undecided at this point, even though it shows the potential of \gr
astronomy of clusters. This is true for both GLAST and ground-based
detectors.


{\bf Starburst Galaxies} presumably harbor many young nonthermal particle
sources collectively seen from a distance. And despite the high star
formation activity the physics should be similar to that in the Milky Way
producing the Cosmic Rays and a large halo. From such a point of view it
would be important to detect nearby objects of this kind in high energy
\grs, such as M82 (at $\sim 3.2$~Mpc) in the Northern, or NGC253 (at $\sim
2.5$~ Mpc) in the Southern Hemisphere \cite{Voel96}. Past attempts have
met mixed success. Despite a long duration observation of about 40 hours,
HEGRA was unable to detect M82. Recently however, CANGAROO-II has
announced a detection of NGC253 at the $11\sigma$ level \cite{Itoh02} . In
contrast to the strictly nuclear starburst of M82, NGC253 should be
extended ($\sim 0.5^{\circ}$ even in \grs, and the full H.E.S.S. I and
CANGAROO-III arrays will attempt to study this object in detail.


The serendipitous discovery of the source TeV J2032+4130 in Cygnus has
provoked the obvious question as to how many such objects are still hidden
in the archives of the past generation of ground-based detectors -- EGRET
had essentially performed such a sky survey from the very beginning. For a
stereoscopic system, where the direction of the primary \gr is determined
on an event by event basis, such a systematic search has been recently
carried out \cite{Pueh03}. Excluding previously detected sources, the
significance distribution per search bin at $E_{\gamma}>500$~GeV is
ssentially Gaussian. A number of weak candidate sources exists which turn
out to be lined up in the Galactic Plane. These are obvious candidates for
future observation with the H.E.S.S. array notwithstanding the conclusion
that nothing significant in {\bf HEGRA's \gr sky} has been overlooked. On
the other hand, only $3.5 \%$ of the total sky had been covered, leaving
lots of room for surprises.

\section{COSMIC RAY ORIGIN}

The question has been one of the prime motivations for \gr astronomy to
begin with. This concerns the population up to a few PeV and, possibly, to
the "ankle" beyond which an extragalactic or even a top down decay origin
seems most likely (an area strictly reserved for air shower physics, see
\cite{Cron03}). Energetically the Galactic population of SNRs appears as the
default solution and this has developed into a folklore in the wider
community. Sociologically it is understandable to circumvent one of the
long standing physics problems of the last century in this form,
scientifically it is not.

Only most recently a combination of SNR observations in hard (synchrotron)  
X-rays and TeV \grs on the one hand, and theory of diffusive shock
acceleration on the other, has brought the Cosmic Ray origin question
close to a physics solution. This development has been summarized in
\cite{Voe03a}. From the theory side this involves the solution of
Fokker-Planck-type kinetic transport equations for nuclei and electrons,
nonlinearly coupled to the hydrodynamics of the thermal gas via the
pressure gradient of the nonthermal component \cite{Bere02,Ber03a}.
\begin{figure}
\centering
\includegraphics[width=7.5cm]{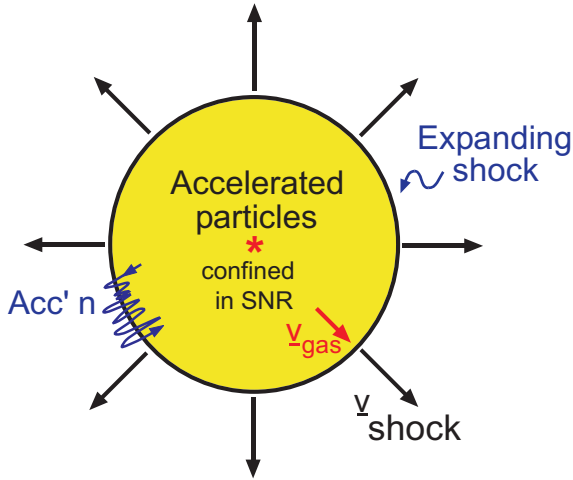}
\caption{Schematic of diffusive shock acceleration at a Supernova 
Remnant.} 
\label{f7}
\end{figure}
Nuclei are only {\it injected} at the quasiparallel parts of the shock
surface, leading also to a dipolar morphology for a uniform external
field, and requiring a renormalization of the spherically symmetric
solution \cite{Voe03b}. At least for young SNRs particle scattering
occurs at the Bohm limit in an amplified effective field $B_{\mathrm eff}$
\cite{Bell01,Bere02}, whereas for older remnants it has been argued that
high energy particles might escape rather effectively \cite{Ptus03} which
would make TeV observations difficult. Since $B_{\mathrm eff}$ and the
injection rate are only calculable within factors of order unity,
one can empirically derive their values from the nonlinear character of
the observed electron synchrotron spectrum. As a result the well-known SNR
SN~1006, assumed to be a SN type Ia as the result of the thermonuclear
explosion of an accreting White Dwarf star, can be consistently
interpreted in terms of a dominant nuclear Cosmic Ray component
accelerated by this object \cite{Bere02,Ber03b}.

Applying this theory to a different object, I will discuss the case of
Cassiopeia A (Cas~A), the youngest known SNR in the Galaxy from around the
year 1680 A.D., and one of the three SNRs claimed to be detected in TeV \grs
\cite{Tani98,Mura00,Aha01b}. Cas~A is the brightest radio source  in the 
Galaxy and amongst
the best-studied objects in the sky, also in other wavelengths. In clear
contrast to SN 1006, Cas~A is believed to be the result of gravitational
collapse of the core of a very massive progenitor, probably a Wolf-Rayet star
with a complex sequence of mass-loss phases Blue Supergiant $\Rightarrow$~Red
Supergiant (RSG) $\Rightarrow$~Wolf-Rayet star $\Rightarrow$~SN~explosion. The
interaction of the fast Wolf-Rayet wind with the massive slow RSG wind has
presumably created a dense shell with mean density of about $10~{\mathrm
 cm}^{-3}$ whose shocked configuration is identified with the bright ring 
in
the radio synchrotron image (Fig. 8), whereas the present SNR shock is assumed
to be already propagating through the unperturbed RSG wind region
\cite{Bork96}.
\begin{figure}
\centering
\vspace*{-5cm}
\includegraphics[width=8.5cm]{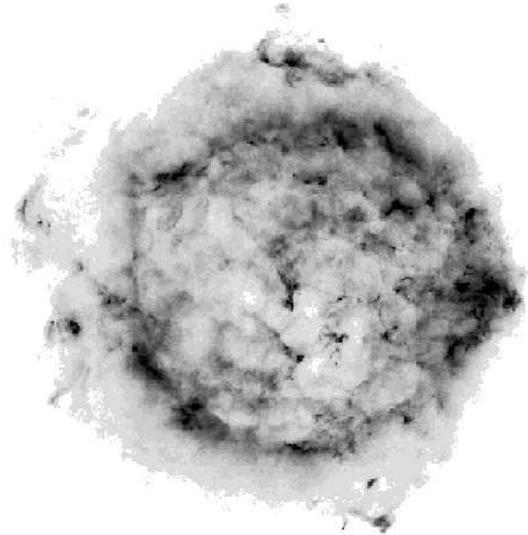}
\caption{Cas~A in 6 cm radio emission, observed with the VLA in 1986 
(Courtesy R.J. Tuffs).} \label{f8}
\end{figure}
The HEGRA system detected Cas~A as a weak source in TeV \grs at 3.3 \% of the
Crab flux \cite{Aha01b}. The numerical solution of the dynamic equations for
the SNR evolution is consistent with a total mechanical SNR energy $E_{\mathrm
SN}= 4\times 10^{50}$~erg as well as an extremely high internal field
$B_{\mathrm eff}=200\mu$~G downstream of the shock, and of order 1 mG in 
the
shell \cite{Ber03a}. For comparison, an estimate from the X-ray morphology
yields about 0.1 mG \cite{Vink03}.

The spatially integrated overall momentum spectrum of accelerated protons
hardens towards the upper cutoff at a $ \mathrm{few}\times 10^{14}$~eV,
whereas the electrons are subject to synchrotron cooling with a cutoff at
about $10^{13}$~eV. This is clearly visible in the derived synchrotron
emission spectrum \cite{Ber03a} whose nonlinear convex shape flattens
around $10^{13}$~Hz and thus stays below the infrared point \cite{Tuff97}
which should contain a thermal dust emission contribution (Fig. 9). The
reasonable fit to the data constitutes at the same time a necessary
condition for the consistency of the theoretical description.
\begin{figure}
\centering
\includegraphics[width=7.5cm]{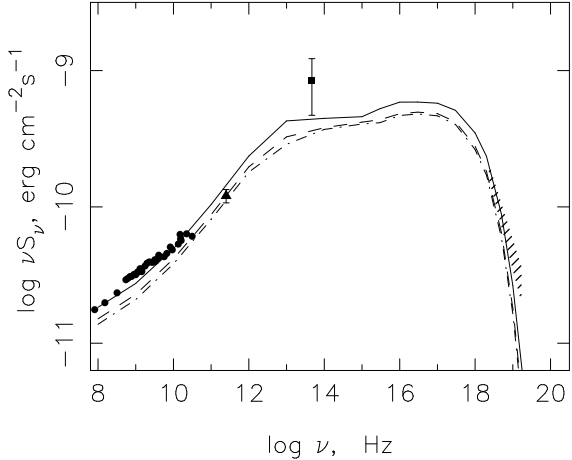}
\caption{Radio and X-ray emission from Cas~A, observed in different years. 
The model curves correspond to the epochs 1970, 2003 and 2022  
\cite{Ber03a}.} \label{f9}
\end{figure}

The shock propagation in a medium of decreasing density is demonstrated by
the secular decline of the synchrotron flux that is clearly visible in
theory and experiment (Fig. 12). In a uniform environment this flux should
instead increase momotonically with time.


\begin{figure}
\centering
\includegraphics[width=7.5cm]{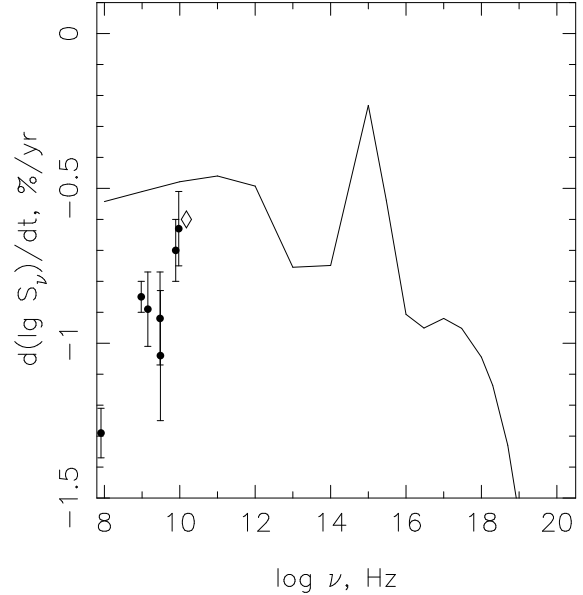}
\caption{Secular decline of the synchrotron flux from Cas~A, in percent 
per yr  
\cite{Ber03a}.} \label{f10}
\end{figure}

The high gas density and effective B-field finally lead to a hadronic \gr
energy flux that exceeds the Bremsstrahlung (NB) and Inverse Compton (IC)  
fluxes by almost two orders of magnitude at 1 TeV, while remaining clearly
below the EGRET upper limit at 100 GeV. This implies a renormalization by
a factor 1/6 \cite{Voe03b}. The conclusion is that Cas~A is a hadronic \gr
source with a strong dominance of accelerated nuclear Cosmic Rays over the
electron component.


\begin{figure}
\centering
\includegraphics[width=7.5cm]{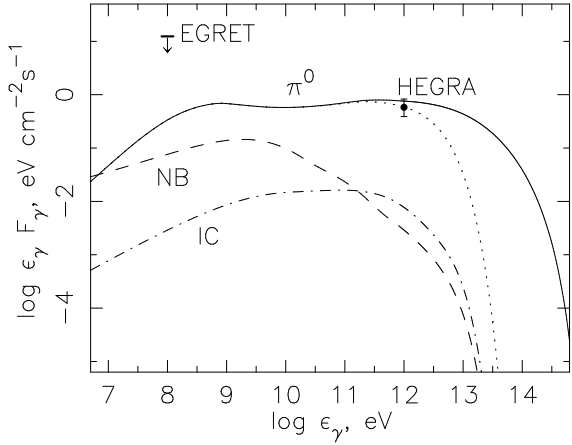}
\caption{Predicted \gr energy flux from Cas~A
\cite{Ber03a}.} \label{f11}
\end{figure}

Obviously Cas~A is a prominent member of the SNR population. At the same
time it is clear that an unambiguous model prediction for its \gr flux
requires a detailed knowledge of the source which can only be obtained
by extensive astronomical observations in other wavelength ranges. For
these reasons an extrapolation from the two cases of SN~1006 and Cas~A
to the effect of the Galactic SNR population is not without risk. Future
observations in high energy \grs and other wavelength ranges, especially
of the "diffuse"  \gr background in the Galactic disk, are desirable to
obtain a statistically impeccable solution.

\section{PERSPECTIVES}

In late 2003 the full imaging array H.E.S.S. I, and probably also
CANGAROO-III, will go into operations. They will do everything faster and
better. But in my view they should not start collecting stamps! We should
rather (i) confront theory of Cosmic Ray origin with detailed \gr spectra
and morphologies and increase the source statistics (ii) attempt to
determine the spectrum of the diffuse \gr background in the Galactic plane
(iii) give Pair Halos a try, not only AGNs, and determine the CIB up to $z
\sim 0.5$, (iv) detect rich clusters of galaxies (v) start an determined
search for Dark Matter in the center of the Galaxy and other nearby Dark
Matter Halos, in complementarity to LHC up to about 10 TeV, and alone
beyond.

GLAST is scheduled to start operations still in 2006. With its large 
field
of view its sky survey will be unsurpassed in the GeV range. GLAST should
be able to identify the majority of the EGRET sources and to investigate
them physically. As regards Cosmic Ray propagation in the
Galaxy, passive \gr sources will play a large role following the tradition
of satellite \gr astronomy. Its is not obvious how much will be learned
from the expected dramatic increase of the AGN population. But clusters of
galaxies may become detectable, and GLAST is expected to study their
morphology in an attempt to find recent accretion events. A complementary
program with ground-based arrays suggests itself.

The aim of a next generation imaging Cherenkov array must be a threshold
of a few GeV, the theoretical limit for this technique. This implies
energies in the satellite range with effective detector areas exceeding
$10^4~{\mathrm m}^2$. A well-known idea is the Atacama project 5@5~(= 5
GeV threshold energy at 5 km altitude) at the ALMA site in Chile. In a
detailed Monte Carlo study \cite{Aha01c} for such a stereo system of 5
large ($600 {\mathrm m}^2$) imaging Cherenkov telescopes an important
signal/noise improvement has been confirmed which is the result of being
closer to the shower maximum at this height. The size and complexity of
such an array suggest a worldwide collaboration beyond the European
institutions involved in the present design studies. Fig. 12 shows a
possible configuration. Such an instrument will be complementary to high
energy arrays like H.E.S.S. Another idea is ECO (= European Cherenkov
Observatory) on La Palma at 2100 m a.s.l., emphasizing even larger mirrors
and employing fast optical detectors with improved quantum efficiency that
are talked about since many years. Monte Carlo studies are under way
\cite{Merc03}.


\begin{figure}
\centering
\includegraphics[width=7.5cm]{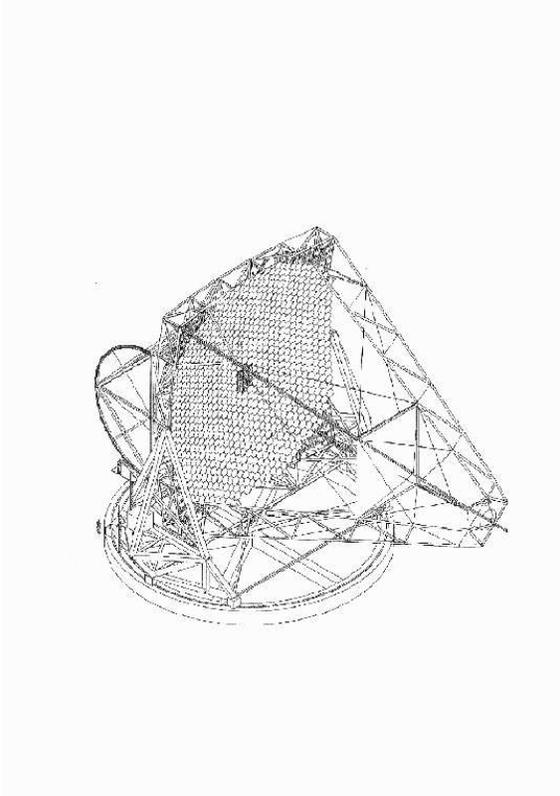}
\caption{Configuration of a $600 {\mathrm m}^2$ Cherenkov telescope, with
steel structure and parabolic mirror support, for a 5 km altitude array.
Not yet adapted to high altitude conditions.}
\label{f12}
\end{figure}

\end{document}